\documentclass[twocolumn,epjc3]{svjour3}
\usepackage{amsmath}

\RequirePackage{fix-cm}
\smartqed
\RequirePackage{graphicx}
\journalname{Eur. Phys. J. C}

\begin{document}

\title{A different interpretation of "Measuring propagation speed of Coulomb
fields" by R. de Sangro, G. Finocchiaro, P. Patteri, M. Piccolo, G. Pizzella}
\author{Anatoly Shabad}
\institute{
 P. N. Lebedev Physics Institute, Leninsky Prospekt 53, Moscow,
117924, Russia
                \and  Tomsk State University, Lenin Prospekt 36, Tomsk 634050,
    Russia.   \\\email{shabad@lpi.ru}}
\date{Received: date / Accepted: date}
\maketitle

\begin{abstract}
We claim that the anti-relativistic statement in Ref. \cite{experiment} that
the Coulomb field of a moving charge propagates rigidly with it, cannot as a
matter of fact be inferred from the measurements reported in that reference.
Registered is not the passing of the Coulomb disk, but the
acceleration-dependent part of the Li\'{e}nard-Wiechert field.

\end{abstract}


\institute{
 P. N. Lebedev Physics Institute, Leninsky Prospekt 53, Moscow,
117924, Russia
                \and  Tomsk State University, Lenin Prospekt 36, Tomsk 634050,
    Russia.   \\\email{shabad@lpi.ru}}

\institute{
 P. N. Lebedev Physics Institute, Leninsky Prospekt 53, Moscow,
117924, Russia
                \and  Tomsk State University, Lenin Prospekt 36, Tomsk 634050,
    Russia.   \\\email{shabad@lpi.ru}}

\institute{
 P. N. Lebedev Physics Institute, Leninsky Prospekt 53, Moscow,
117924, Russia
                \and  Tomsk State University, Lenin Prospekt 36, Tomsk 634050,
    Russia.   \\\email{shabad@lpi.ru}}

\subtitle{No rigid propagation of Coulomb field}

\titlerunning{No rigid propagation}

\authorrunning{A.E.Shabad}


\bigskip

According to the generally accepted Feynman's view \cite{Feynman} a certain
time is needed for a charge to gain its disk-like shape prescribed by the Li%
\'{e}nard-Wiechert formulas \cite{Landau2,Jackson} \footnote{%
It was explicitly demonstrated in \cite{GitShaShi} that these formulas and
the one given in \cite{Feynman} are identical} after this charge starts its
homogeneous motion (see \cite{Feinberg}). In contrast with this view fully
respecting the finiteness of the speed of propagation of interaction, the
experimental results \cite{experiment} obtained in Frascati National
Laboratory on measuring the electric field of a charge injected into the
working space in a certain point and then left to move freely with the speed
$v$ of the order of that of light, $v=c(1-0.5\cdot 10^{-6}),$ are
interpreted by their authors as witnessing in favour of instantaneous
propagation. This interpretation already called into being some speculations
\cite{Field} about superluminal effects and causality violation. We are
going to explain here that this interpretation can be avoided by taking into
account the acceleration phase of the charge.

Let us imagine first that the charge is point-like and that it is emitted in
the point $z^{\prime }=0,$ $y^{\prime }=0$ and then moves along the axis $z$
following the law%
\begin{equation}
z^{\prime }=vt^{\prime },\text{ }y^{\prime }=0,  \label{law}
\end{equation}%
with $z^{\prime }$ and $y^{\prime }=0$ being the coordinates of its position
at the time instance $t^{\prime }.$ The field is observed at the
time-instance $t$ in the point $z,$ $y$ provided that it has been created by
the charge when it was in the point $z^{\prime }$ and $y^{\prime }=0$ at the
time $t^{\prime }$ if the "light cone" equation
\begin{equation}
\left( z-z^{\prime }\right) ^{2}+y^{2}=c^{2}\left( t-t^{\prime }\right) ^{2},
\label{light cone}
\end{equation}%
is obeyed, since the electromagnetic interaction propagates exactly with the
speed of light $c.$

The important circumstance necessary to understand the experiment,
rightfully emphasized by the authors, is that the disk accompanying the fast
moving charge is so narrow that the field can be registered only at its
maximum, otherwise it is many orders of magnitude smaller and beyond the
experimental sensitivity. The maximum of the field passes the point $z,$ $y$
at the observation time
\begin{equation}
t=\frac{z}{v},  \label{maxtime}
\end{equation}%
as it follows from the form of the dominator, $\gamma =\left( 1-\frac{v^{2}}{%
c^{2}}\right) ^{-1/2}$%
\begin{equation}
R^{\ast 3}=\left[ \left( z-vt\right) ^{2}+\frac{y^{2}}{\gamma ^{2}}\right] ^{%
\frac{3}{2}}  \label{denominator}
\end{equation}%
in the expression for the electric field of a moving charge, see \ Eq.
(38,6) in \cite{Landau2} and also Eq. (\ref{L-W}) below. Using (\ref{law})
and (\ref{maxtime}) we obtain Eq. (6) in Ref. \cite{experiment}%
\begin{equation}
t^{\prime }=t-\frac{y}{c}\gamma  \label{t'}
\end{equation}%
as solution to equation (\ref{light cone}) subject to the retardation
condition $t^{\prime }<t.$

In the experiment, the sensors were located in the points, whose coordinate $%
z$ varied from 1 to 5 m, and $y$ from 3 to 55 cm. With these values
substituted into (\ref{t'}), (\ref{maxtime}) one gets the vast negative
value for the time difference $t^{\prime }-t$ (up to $-1800$ ns$)$, which
means that the point $z^{\prime }=z-y\gamma $ where the registered field has
been created, is separated from the observation point by the tremendous
negative distance up to $z^{\prime }=-y\gamma =-550$ m $(z\ll y\gamma ).$
However, the distance of hundred meters for the beam to move before it is
registered is not available in the experiment. In other words, it comes out
that the field had been paradoxically created long before its source
appeared. In short, we face a "disproof" of Special Relativity by \textit{%
reductio ad absurdum,} because the above consideration is based on its
postulates. The authors of Ref. \cite{experiment} suggest to resolve the
paradox by concluding that the charge appears with its disk field already
formed, in other words, that the Coulomb field propagates rigidly together
with the charge carrying it. This conclusion is, however, anti-relativistic
in itself, since it admits forming the field in the whole space at once
\footnote{%
The natural view \cite{Feinberg} is that only the microscopic core of the
charge, where its field mass is mostly gained (for instance, due to its
nonlinear self-interaction \cite{Caio}) is an integral part of it and thus
may be thought of as ever accompanying it. As long as the electron is
concerned, this is its classical radius of the order of 3 fm. Certainly, the
microcausality is questioned within this assumption. However, the present
measurements do not deal with such small distances.}.

To refute this disproof, it is necessary to consider a fuller problem of the
charge being accelerated before it gets into the registration space
(experiment hall). Under the acceleration, either the speeding up of the
beam or its magnetic bending may be understood.

The full Li\'{e}nard-Wiechert formula \cite{Landau2,Jackson} for accelerated
charge is%
\begin{eqnarray}
\mathbf{E(r,}t) &=&e\frac{1-\frac{v^{2}}{c^{2}}}{\left( R-\frac{\mathbf{Rv}}{%
c}\right) ^{3}}\left( \mathbf{R}-\frac{\mathbf{v}}{c}R\right) +  \label{L-W}
\\
&&+e\frac{\left[ \mathbf{R\times }\left[ \left( \mathbf{R}-\frac{\mathbf{v}}{%
c}R\right) \times \frac{\overset{\cdot }{\mathbf{v}}}{c^{2}}\right] \right]
}{\left( R-\frac{\mathbf{Rv}}{c}\right) ^{3}},
\end{eqnarray}%
where $\overset{\cdot }{\mathbf{v}}=\frac{d\mathbf{v}}{dt^{\prime }}$ is the
acceleration of the charge at the moment when it created the electromagnetic
field to be registered at the time $t$ in the point $(z,$ $y).$The vector $%
\mathbf{R=}\left( z-z^{\prime },\text{ }y\right) $ is drawn from the point $%
\left( z^{\prime },0\right) $, where the field is created to the observation
point $\left( z,y\right) $. Its modulus $|\mathbf{R}|=R=c(t-t^{\prime })$ is
the distance between these points in agreement with (\ref{light cone}). The
identity of $R^{\ast 3}$ (\ref{denominator}) with the denominators in (\ref%
{L-W}) is seen taking into account that $\mathbf{Rv}=v(z-z^{\prime }).$ Then%
\begin{equation}
R-\frac{\mathbf{Rv}}{c}=c(t-t^{\prime })-\frac{v}{c}(z-z^{\prime }),  \notag
\end{equation}%
which coincides with $R^{\ast }$ after the substitution of the light-cone
condition (\ref{light cone}) and the charge trajectory (\ref{law}) into (\ref%
{denominator}). The vector $\mathbf{R}-\frac{\mathbf{v}}{c}R$ $=$ $\mathbf{R}%
-\mathbf{v}(t-t^{\prime })=$

=$\left( z-z^{\prime }-v(t-t^{\prime }),\text{ }y\right) =$ $\left( z-vt,%
\text{ }y\right) $ in the numerators of (\ref{L-W}) is the vector drawn from
the point $\left( vt,\text{ }0\right) $ where the charge is \footnote{%
to be more precise, where it would be if it continues to move with the same
constant speed after it has created the observed field.} at the moment of
observation to the observation point $\left( z,\text{ }y\right) $.

The first term in (\ref{L-W}) is suppressed by the factor $\gamma ^{-2}=1-%
\frac{v^{2}}{c^{2}}=10^{-6}.$ (This fact has forced the authors of Ref. \cite%
{experiment} to admit that the field of the moving charge might be
registered only in the maximum point \ref{maxtime}). On the contrary, the
second term does not contain this small factor. To estimate its value let us
assume that the beam has been turned by the bending magnets short before the
injection as it is characteristic of the Beam Test Facility in the Frascati
National Laboratory, and that the curvature radius is of the same centimeter
scale as $R.$ Then the centrifugal acceleration is $|\overset{\cdot }{%
\mathbf{v}}|=\frac{v^{2}}{R},$ and the extra factor in the second term in (%
\ref{L-W}) as compared with the first term is of the order of unity: \ $R%
\frac{\overset{\cdot }{v}}{c^{2}}=R\frac{1}{c^{2}}\frac{v^{2}}{R}\approx 1,$
not suppressed by the factor $10^{-6}$ present in the first term. This may
signify that it is the field given by the second, radiative term in (\ref%
{L-W}) that is registered in the experiment, and, moreover, not in the point
of its maximum, but when it first appears, being created at the time of
acceleration. If, quite roughly, we imagine, following Ref. \cite{Feinberg},
that the charge was immediately accelerated in the point $z^{\prime
}=y^{\prime }=0$ at the time instance $t^{\prime }=0,$ the spherical wave is
created propagating with the speed of light $c.$ It reaches the sensors
located at the points $z=z_{1},z_{2},$ $y=$ $y_{1},y_{2}$ at times $%
t=t_{1},t_{2},$ respectively, with $%
c^{2}t_{1,2}^{2}=z_{1,2}^{2}+y_{1,2}^{2}. $ The observed longitudinal
"speed" $V_{\text{long}}$ of the registered signal between these points
defined as (we mean $z_{1}<z_{2},$ $t_{1}<t_{2},$) the distance along the
z-axis $\left( z_{2}-z_{1}\right) $ divided by the difference of the times $%
\left( t_{2}-t_{1}\right) ,$\ at which the wave reaches the two sensors, is
\begin{equation}
V_{\text{long}}=\frac{c\left( z_{2}-z_{1}\right) }{\left(
y_{2}^{2}+z_{2}^{2}\right) ^{\frac{1}{2}}-\left( y_{1}^{2}+z_{1}^{2}\right)
^{\frac{1}{2}}}.  \label{Vlong}
\end{equation}%
This is not determined by the speed of the beam, which is ever smaller than $%
c.$ Contrary to the latter, $V_{\text{long}}$ may be larger than $c$
(without contradicting principles of Special Relativity, since no
information can be transmitted with this "speed"). Besides, $V_{\text{long}}$
is not a universal, but depends upon the choice of the sensor positions. If $%
y_{1}=y_{2}=y$ the following chain of inequalities, obtained each by
squaring the previous one%
\begin{eqnarray}
z_{2}-z_{1} &>&\left( y^{2}+z_{2}^{2}\right) ^{\frac{1}{2}}-\left(
y^{2}+z_{1}^{2}\right) ^{\frac{1}{2}},  \label{y} \\
\left( y^{2}+z_{2}^{2}\right) ^{\frac{1}{2}}\left( y^{2}+z_{1}^{2}\right) ^{%
\frac{1}{2}} &>&y^{2}+z_{1}z_{2},  \notag \\
z_{2}^{2}+z_{1}^{2} &>&2z_{1}z_{2},\text{ \ \ \ \ \ \ \ \ }  \notag
\end{eqnarray}%
proves that $V_{\text{long}}>c$ in such case. If next we go to larger values
of $y_{1},$ $y_{1}>y_{2}$, the inequality $z_{2}-z_{1}>\left(
y_{2}^{2}+z_{2}^{2}\right) ^{\frac{1}{2}}-\left( y_{1}^{2}+z_{1}^{2}\right)
^{\frac{1}{2}}$ becomes even stronger than what it was when $y_{1}=y_{2}$, (%
\ref{y}). The "speed" $V_{\text{long}}$ becomes larger and larger, till $%
y_{1}$ reaches the root of the equation\ $\left( y_{2}^{2}+z_{2}^{2}\right)
^{\frac{1}{2}}-\left( y_{1}^{2}+z_{1}^{2}\right) ^{\frac{1}{2}}=0,$ where
the denominator in (\ref{Vlong}) turns to zero. In this case the wave comes
to the both sensors simultaneously, hence it is no wonder that $V_{\text{long%
}}=\infty .$ So, we conclude that $V_{\text{long}}>c$ at least as long as $%
y_{1}\geq y_{2}.$

Let us now turn to the results of measuring the longitudinal "speed" as
these are presented in Table 1 of Ref. \cite{experiment}. The first remark
concerning this Table is that even taking into account the indeterminacies
stemming from the finite sizes of the sensors one cannot state -- contrary
to what the authors do -- that the speeds between different couples of
sensors are one speed, as these should have been if the speed were that of
the beam. The second remark is that the latter must be smaller than (almost
equal to) $c$, whereas the results listed in Table 1 are at least in certain
cases definitely above the speed of light. Unfortunately, the values of the
y-positions of the sensors, between which the speeds were measured, are not
indicated by the authors, and this fact makes it impossible to imply the
model formulas (\ref{Vlong}). Nevertheless, if we calculate the speed
following the data given in the middle column of the first line of Table 1
we get \bigskip
\begin{equation*}
V_{\text{long}}=\frac{223.0\pm 1.5}{7.28\pm 0.02}\frac{cm}{ns}.
\end{equation*}%
This ranges within $(3.092\div 3.033)\cdot 10^{10}cm/s,$ which is wholly and
essentially above the speed of light. The value $V_{\text{long}}=$ $%
3.063\cdot 10^{10}cm/s$ supplied by Eq. (\ref{Vlong}) with the values $%
z_{1}=329.5cm,$ $z_{2}=552.5cm,$ $y_{1}=55cm,$ $y_{2}=5cm$ fits this range.

To be convinced that they are really registering the coming of the Coulomb
field to their sensors, the authors \ of \cite{experiment} performed the
background measurements, when a filter of lead was placed in the way of the
charge before it might reach the projection points $z$ of the sensors. The
results presented in their Fig. 15 indicate that the countings are much
smaller in that case. However, this fact cannot be taken as a proof that
really the passing of the disk is registered when there is no filter,
because after the Coulomb field is formed at time $t^{\prime }$ it continues
to exist afterwards irrespective of the consequent fate of the charge, i.e.
even after it is absorbed by the filter. The results of the background
measurements can be only understood as an indication that not solely the
charge is absorbed by the lead filter, but also its Coulomb field, as well
as the radiation field. An indirect confirmation to this assumption may be
in that the counting in Fig. 15 seem somewhat growing with the growth of the
transversal position of the sensor $y,$ when the admitted screening may be
expected to become less efficient. This assumption is favoured by the fact
that the counting for the sensor most remote from the beam axis, $y=55cm,$
is practically the same irrespective of whether the filter is present.

Our conclusion is that most probably the signal registered in the Frascati
experiment \cite{experiment} originates from the radiation due to the
acceleration of the beam, and does not belong to the Coulomb disk of the
charge. As for the latter, it should be sought for hundreds of meters ahead,
already outside of the laboratory, after the beam itself is absorbed by its
concrete wall, unless, certainly, it is screened by other possible objects.

\bigskip
\begin{acknowledgements}
Supported by RFBR under Project 14-02-01171, and by the TSU
Competitiveness Improvement Program, by a grant from
\textquotedblleft The Tomsk State University D.I. Mendeleev
Foundation Program\textquotedblright. \end{acknowledgements}


\begin{thebibliography}{9}
\bibitem{experiment} R. de Sangro, G. Finocchiaro, P. Patteri, M. Piccollo,
and G. Pizzella, Measuring Propagation Speed of Coulomb Fields, Eur. Phys.
J. C \textbf{75}, 137 (2015).

\bibitem{Feynman} R.P. Feynman, R.B. Leighton and M. Sands, The Feynman
Lectures in Physics, "Electromagnetism II" Ch 21-1. Addison-Wesley, Reading
Massachusetts, (1963). See also
http://www.feynmanlectures.caltech.edu/II\_21.html\#Ch21-S3.

\bibitem{Landau2} L.D. Landau, E.M. Lifshitz, The Classical Theory of
Fields, 509p. GIF-ML, Moscow (1962); Pergamon Press, Oxford (1971).

\bibitem{Jackson} J.D. Jackson, Classical Electrodynamics, pp. 654 -- 658.
Wiley, New York, (1962 -- 1975).

\bibitem{GitShaShi} D.M. Gitman, A.E. Shabad and A.A. Shishmarev, "Note on
"Measuring Propagation Speed of Coulomb Fields" by R. de Sangro, G.
Finocchiaro, P. Patteri, M. Piccollo, and G. Pizzella", arXiv:1605.02545
[physics.class-ph]\ (2016), Eur. Phys. J. C 76(5) April 2016

\bibitem{Feinberg} E.L. Feinberg, Zh. Eksp. Teor. Fiz., \textbf{50}, 202
(1966); E.L. Feinberg in: Problems of Theoretical Physics, A Memorial Volume
to Igor Tamm, pp. 248 -- 264 (NAUKA, Moscow, 1972) -- in Russian

\bibitem{Field} J.H. Field, Comment on "Measuring propagation speed of
Coulomb fields", arXiv:1506.06630[physics.gen-ph] (2015) (or
arXiv:1506.06630v2 [physics.gen-ph] (2016) for completed version).

\bibitem{Caio} C.V. Costa, D.M. Gitman and A.E. Shabad, Phys. Scr. \textbf{90%
}, 074012 (2015); D.M. Gitman, A.E. Shabad and A.A. Shishmarev, "Moving
point charge in nonlinear electrodynamics", arXiv:1509.06401v2[hep-th]
(2015).
\end{thebibliography}
\end{document}